\documentclass[trackchanges]{aastex701}

\begin{document}
\title{Magnetic field configuration favoring X-class solar flares: violation of Joy's and Hale's laws }

\author[0000-0002-1396-7603]{Liu, S.}
\affiliation{State Key Laboratory of Solar Activity and Space Weather, National Astronomical Observatories, Chinese Academy of Sciences, Beijing 100101, China}
\email{lius@nao.cas.cn}

\author[0000-0002-5152-7318]{Su, J.T.}
\affiliation{State Key Laboratory of Solar Activity and Space Weather, National Astronomical Observatories, Chinese Academy of Sciences, Beijing 100101, China}
\affiliation{School of Astronomy and Space Sciences, University of Chinese Academy of Sciences, Beijing 100049, China}
\email{sjti@nao.cas.cn}

\author[0000-0002-3141-747X]{Zhang, M.}
\affiliation{State Key Laboratory of Solar Activity and Space Weather, National Astronomical Observatories, Chinese Academy of Sciences, Beijing 100101, China}
\affiliation{School of Astronomy and Space Sciences, University of Chinese Academy of Sciences, Beijing 100049, China}
\email{zhangmei@nao.cas.cn}

\author[0000-0003-2544-9544]{Wang, J.X.}
\affiliation{State Key Laboratory of Solar Activity and Space Weather, National Astronomical Observatories, Chinese Academy of Sciences, Beijing 100101, China}
\affiliation{School of Astronomy and Space Sciences, University of Chinese Academy of Sciences, Beijing 100049, China}
\email{wangjx@nao.cas.cn}

%\author[0000-0001-5002-0577]{Jiang, J.}
%\affiliation{School of Space and Earth Sciences, Beihang University, Beijing, China}
%\affiliation{Key Laboratory of Space Environment Monitoring and Information Processing of MIIT, Beijing, China}
%\email{jiejiang@buaa.edu.cn}

\begin{abstract}
We investigate the statistical relationship between magnetic tilt angle and X-class flare productivity using {\bf 39} flare-productive active regions that produced {\bf 64} X-class flares. By classifying global magnetic tilt properties into four quadrants based on compliance with Hale's and Joy's laws, we identify their distinctive flare productivity characteristics. The Green quadrant (following both laws) contains {\bf 18} active regions producing {\bf 24} X-class flares, while the Yellow quadrant (following Hale's law but violating Joy's law) shows significantly enhanced productivity with {\bf 18} active regions producing {\bf 36} X-class flares. The Black quadrant (violating both laws), though containing only 3 active regions, produces 4 X-class flares. More importantly, flares in both the Yellow and Black quadrants exhibit systematically higher flare classes compared to those in the Green quadrant. Our statistical analysis demonstrates that violation of Hale's or Joy's law at the global scale are strongly associated with increased flare occurrence and flare intensity. Furthermore, examination of globally normal (Green quadrant) regions reveals that localized tilt anomalies are universally present and also contribute to X-class flare production. These results establish that abnormal magnetic tilt configurations — whether occurring at the global or local scale — are key indicators of major flare activity.
\end{abstract}


%%%%%%%%%%%%%%%

\section{Introduction} 
\label{sec:intro}

Solar flares, among the most powerful explosive phenomena in the solar system, originate from the sudden release of magnetic non-potential energy stored in active regions (ARs) \citep{2011LRSP....8....6S, 2011SSRv..159...19F, 2017LRSP...14....2B}. Accurate prediction of such events—particularly the most disruptive X-class flares—remains a fundamental challenge in solar physics and space weather forecasting. Addressing this challenge requires identifying reliable magnetic field parameters that can effectively characterize an active region's eruptive potential.

The configuration and evolution of sunspot magnetic fields are governed by fundamental laws. The Hale's polarity law describes the systematic orientation of leading and following polarities in bipolar ARs, which reverses between the northern and southern hemispheres and from one 11-year solar cycle to the next, defining the 22-year magnetic cycle \citep{1919ApJ....49..153H, 1925ApJ....62..270H}. Concurrently, the Joy's law states that the magnetic axis of a bipolar group is typically tilted, with the leading polarity spot closer to the equator, and this tilt increases with the latitude \citep{1988assu.book.....Z}. These empirical laws reflect the large-scale organization of solar dynamo and the influence of Coriolis forces on the rising magnetic flux tubes.

Generally, the majority of active regions possess typical magnetic configurations following the established Hale's and Joy's laws. However, a small but significant subset displays anomalous properties violating Hale's and Joy's laws. \cite{1948ApJ...107...78R} reported that 3.1\% of bipolar groups observed at Mount Wilson Observatory violated sunspot polarity laws, while \cite{1989SoPh..124...81W} found only 4\% of 2700 surveyed bipolar magnetic regions exhibited reversed east-west polarity orientations. These ``rogue" active regions with anti-Hale and/or anti-Joy configurations demonstrate particularly anomalous behavior, with dynamo models suggesting that even a single such region could significantly influence solar cycle evolution \citep{2017SoPh..292..167N, 2014JGRA..119..680C}. The formation of such abnormal active regions appears linked to complex sunspot configurations, particularly $\delta$-type sunspots where colliding bipolar groups create magnetic structures with polarity arrangements often violating Hale's law \citep{1983SoPh...89...43T}.

The emergence pattern of an active region is intrinsically linked to its adherence to Hale's and Joy's laws, a connection fundamentally governed by the twist and writhe of its magnetic flux tube. Here, the writhe quantitatively represents the large-scale magnetic tilt under Joy's law, serving as a direct measure of the global twist of the field, while both writhe and twist are key topological components of magnetic helicity that can dynamically convert into one another. The relationship between these components is complex and sample-dependent, with studies reporting both positive \citep{1998ASPC..140..131C} and negative \citep{2001A&A...374..294T, 2003A&A...397..305L} correlations, likely influenced by the pre-existing helicity profile when the flux system emerges from the deep convective zone \citep{1992RSPSA.439..411M}. This writhe-twist interplay directly governs the resulting photospheric magnetic configuration, where numerical simulations show how inherent complexity would emerge and manifest as multi-polar structures \citep{1998ApJ...493L..43M, 2014SoPh..289.3351T, 1999ApJ...522.1190L, 1999ApJ...521..460F, 2015ApJ...806...79F}. A prime consequence is the formation of $\delta$-sunspots through intense magnetic winding and flux collision, configurations exceptionally prone to accumulating magnetic helicity and non-potential energy, thereby furnishing a robust foundation for prolific solar eruptive activity.

A significant body of research has established that active regions violating Hale's and/or Joy's laws—denoted as anti-Hale and/or anti-Joy regions—are statistically more flare-productive \citep{2002SoPh..209..361T, 2005SoPh..229...63T, 2015ApJ...809...34C, 2017ApJ...834...56T, 2019ApJ...873...23Z}. Such regions, characterized by anomalous polarity orientations or tilt angles that significantly deviate from Joy's law (i.e., abnormally large or even inverted tilts), typically possess highly sheared magnetic configurations. These non-potential magnetic field structures are direct evidence of strong electric currents and signify the accumulation of substantial magnetic helicity and free magnetic energy available for flares {\bf{\citep{2005ARA&A..43..103Z, 2019ApJ...880L...6T, 2019ApJ...887...64T, 2021ApJ...922...41T, 2022ApJ...937...59G, 2024A&A...686A.148S}}}.

Case studies of individual eruptive regions provide compelling evidences for the critical role of magnetic non-potentiality. A classical example is AR 12673, which produced the X9.3 flare—the most powerful event of Solar Cycle 24. It exhibited localized anti-Hale characteristics, where magnetic flux with anomalous polarity orientation was a key factor in building up the extreme non-potentiality required for such an eruption \citep{2019ApJS..240...11P, 2023ApJ...942...27L}. Another illustrative case is AR 12882, where localized anti-Hale characteristics were manifested through a single satellite sunspot configuration. Its complex magnetic topology created a flare-productive environment that led to significant eruptive activity \citep{2022ApJ...937L..11X}.  Similar satellite sunspot configurations, featuring a mix of small-scale magnetic features around a primary sunspot, also indicate heightened major flare activity \citep{2024ApJ...975...46J}.  A key insight is that, despite the co-presence of both Hale-Joy compliant and Hale-Joy violating features, the eruptive potential maybe primarily governed by those exhibiting anti-Hale and/or anti-Joy characteristics.

The eruptive potential associated with the abnormal evolution of magnetic tilt—systematic deviations from the mean patterns described by Joy's law—has seldom been studied as a standalone phenomenon. Rather, it has often been subsumed within broader statistical analyses of active region properties \citep{1999SoPh..189..305T, 2003ApJ...595.1277L, 2003A&A...407L..13T, 2016ApJ...821..127B, 2021JSWSC..11...39G}, such as those relating magnetic flux, complexity, or shear to flare productivity, without being explicitly isolated or quantitatively addressed as a primary physical driver.

In this study, we introduce a novel quantitative framework that classifies magnetic tilt state of active regions into four distinct ``quadrants", providing a unified diagnostic that simultaneously evaluates compliances with both Hale's and Joy's laws. This approach enables us to categorize ARs based on their magnetic configuration status. Based on this classification we perform a systematic statistical analysis of 64 X-class flares from 39 ARs. The paper is structured as follows: Section 2 describes the data sources and the methodology. Section 3 presents the analysis and the statistical results. Section 4 states the main conclusions and finally brief discussions are also provided.

%%%%%%%%%%%%%%%

\section{Data, Sample and Method}
\label{sec:observation}

\subsection{Magnetic Field Data}

The Helioseismic and Magnetic Imager (HMI) aboard the Solar Dynamics Observatory (SDO) provides full-disk vector magnetic field observations with 0.5$\arcsec$ spatial resolution and 12-minute temporal cadence \citep{2012SoPh..275..229S}. Operating at the Fe I 617.3 nm spectral line, which exhibits high sensitivity to the Zeeman effect, HMI samples intensity, Doppler velocity, and polarization at six wavelength positions to reconstruct the full Stokes vector (I, Q, U, V).

The photospheric vector magnetic field is derived from Stokes profiles using the Very Fast Inversion of the Stokes Vector (VFISV) algorithm, which employs a Milne-Eddington atmospheric model to determine field strength, inclination, and azimuth \citep{2011SoPh..273..267B}. Transverse field components are optimized using a minimum energy approach that simultaneously minimizes the divergence of the magnetic field ($\nabla\cdot\mathbf{B}$) and the current density ($\mathbf{J} = \nabla\times\mathbf{B}$) through a simulated annealing scheme \citep{1994SoPh..155..235M, 2006SoPh..237..267M}.

For this study, we utilize the Space-weather HMI Active Region Patches (SHARP) vector magnetograms of the hmi.sharp\_cea\_720s series \citep{2014SoPh..289.3549B, 2014SoPh..289.3483H}, which have been remapped to the Cylindrical Equal-Area (CEA) coordinate system. The CEA projection minimizes geometric distortion and preserves area, making it particularly suitable for tracking active regions and for calculating accurate magnetic tilt angles. These data provide corrected magnetic field vectors with reliable disambiguation of the 180° azimuth ambiguity, essential for our quadrant classification scheme.

\subsection{GOES X-ray Flux Data}

The Geostationary Operational Environmental Satellite (GOES) system provides continuous monitoring of solar X-ray emissions through its X-ray Sensors (XRS), which measures flux in two wavelength bands: 0.5--4 \AA\ and 1--8 \AA. These observations form the standard for solar flare classification (A, B, C, M, and X classes) and are essential for determining flare timing and magnitude  \citep{1994SoPh..154..275G}.

With a temporal cadence of one minute, GOES X-ray data enable precise identification of flare start, peak, and end times. In this study, we utilize the 1--8 \AA\ band measurements specifically to establish the correlation between flare activity and the evolution of magnetic tilts in active regions, facilitating a systematic investigation of how temporal variations in magnetic configuration relate to flare productivity.

\subsection{The Sample}

This study investigates the relationship between magnetic tilt evolution and flare productivity by analyzing a carefully selected sample of 64 X-class flares from 39 solar active regions (ARs). These ARs produced distinct X-class flares between 2008 and 2025, spanning the complete solar cycle 24 and the ongoing solar cycle 25. The sample was selected from approximately 110 X-class flares recorded during this period based on three stringent criteria: (1) the selected AR must have produced at least one X-class flare; (2) the location of the flare in the selected AR must be away from the solar limbs (particularly the east limb) where reliable photospheric magnetic field measurements for tilt calculations before the flare are available; and (3) vector magnetic field data must be available from the Space-weather HMI Active Region Patches (SHARP) series, which provides optimized inversions with reliable projection corrections.

This selection ensures the data quality necessary for a robust statistical analysis. Importantly, our sampling strategy targeted individual X-class flare events rather than active regions. As a result, Table \ref{tab:flare_summary} lists 64 flares from 39 ARs, with some ARs contributing multiple events (e.g., AR 12192 and AR 13664 each produced six X-class flares). The listed information includes the NOAA AR number, solar cycle number, location (the first letter N or S indicates the northern or southern hemisphere respectively), flare timing, flare class, tilt angle (in degrees), the tilt quadrant classification and sunspot type based on the analysis described in the following.

\begin{table}[htbp]
\centering
\caption{Summary of the sample information}
\label{tab:flare_summary}
\scriptsize
\begin{tabular}{cccccccccc}
\hline
NOAA No. & Cycle & Location & Start Time & Peak Time & End Time & Class & Tilt Angle (degrees) & Tilt Quadrant & Sunspot Type \\
\hline
11158 & 24 & S21E13 & 2011/2/15 1:44 & 1:56 & 2:06 & X2.2 & 196.8 & NHJ & $\beta$$\gamma$/$\beta$$\gamma$ \\
11166 & 24 & N10E11 & 2011/3/9 23:13 & 23:23 & 23:29 & X1.5 & 3.1 & AJ & $\beta$$\gamma$$\delta$ \\
11263 & 24 & N18E74 & 2011/8/9 7:48 & 8:05 & 8:08 & X6.9 & 13.5 & AJ & $\beta$$\gamma$$\delta$ \\
11283 & 24 & N17E11 & 2011/9/6 22:12 & 22:20 & 22:24 & X2.1 & 348.2 & NHJ & $\beta$$\gamma$$\delta$/$\beta$$\gamma$ \\
 & 24 & N17E25 & 2011/9/7 22:32 & 22:38 & 22:44 & X1.8 & 348.3 & NHJ & $\beta$$\gamma$$\delta$/$\beta$$\gamma$ \\
11402 & 24 & N23E87 & 2012/1/27 17:37 & 18:37 & 18:56 & X1.7 & 358.7 & NHJ & $\beta$$\gamma$/$\beta$$\gamma$ \\
11429 & 24 & N18W24 & 2012/3/7 0:02 & 0:24 & 0:40 & X5.4 & 186.9 & AHJ & $\beta$$\gamma$$\delta$ \\
 & 24 & N18W48 & 2012/3/5 2:30 & 4:09 & 4:43 & X1.1 & 199.8 & AHJ & $\beta$$\gamma$$\delta$ \\
11520 & 24 & S21E14 & 2012/7/12 15:37 & 16:49 & 17:30 & X1.4 & 165.3 & AJ & $\beta$$\gamma$$\delta$ \\
11875 & 24 & N09E61 & 2013/10/28 1:41 & 2:03 & 2:12 & X1.0 & 353.5 & NHJ & $\beta$$\gamma$$\delta$/$\beta$$\gamma$$\delta$ \\
11893 & 24 & S16E68 & 2013/11/19 10:14 & 10:26 & 10:34 & X1.0 & 170.9 & AJ & $\beta$$\delta$ \\
11944 & 24 & S13W10 & 2014/1/7 18:04 & 18:32 & 18:58 & X1.2 & 186.8 & NHJ & $\beta$$\gamma$$\delta$/$\beta$$\gamma$$\delta$ \\
12017 & 24 & N07E30 & 2014/3/29 17:35 & 17:48 & 17:54 & X1.0 & 345.4 & NHJ & $\beta$$\delta$/$\beta$ \\
12158 & 24 & N15W06 & 2014/9/10 17:21 & 17:45 & 18:20 & X1.6 & 247.0 & AHJ & $\beta$$\gamma$$\delta$ \\
12192 & 24 & S13E37 & 2014/10/26 10:04 & 10:56 & 11:18 & X2.0 & 179.3 & AJ & $\beta$$\gamma$$\delta$ \\
 & 24 & S13W15 & 2014/10/22 14:02 & 14:28 & 14:50 & X1.6 & 178.0 & AJ & $\beta$$\gamma$$\delta$ \\
 & 24 & S13E53 & 2014/10/27 14:12 & 14:47 & 15:09 & X2.0 & 180.0 & AJ & $\beta$$\gamma$$\delta$ \\
 & 24 & S13E27 & 2014/10/25 16:55 & 17:08 & 18:11 & X1.0 & 179.2 & AJ & $\beta$$\gamma$$\delta$ \\
 & 24 & S13E16 & 2014/10/24 21:07 & 21:41 & 22:13 & X3.1 & 179.7 & AJ & $\beta$$\gamma$$\delta$ \\
 & 24 & S13W61 & 2014/10/19 4:17 & 5:03 & 5:48 & X1.1 & 179.5 & AJ & $\beta$$\gamma$$\delta$ \\
12242 & 24 & S15E46 & 2014/12/20 0:11 & 0:28 & 0:55 & X1.8 & 161.9 & AJ & $\beta$$\gamma$$\delta$ \\
12673 & 24 & N14E18 & 2017/9/6 11:53 & 12:02 & 12:10 & X9.3 & 101.4 & AJ & $\beta$$\gamma$$\delta$ \\
 & 24 & N14E33 & 2017/9/7 14:20 & 14:36 & 14:55 & X1.3 & 98.3 & AJ & $\beta$$\gamma$$\delta$ \\
 & 24 & N14E17 & 2017/9/6 8:57 & 9:10 & 9:17 & X2.2 & 102.6 & AJ & $\beta$$\gamma$$\delta$ \\
12887 & 25 & S26E05 & 2021/10/28 15:17 & 15:35 & 15:48 & X1.0 & 55.5 & NHJ & $\beta$$\gamma$/$\beta$ \\
12975 & 25 & N17E24 & 2022/3/30 17:21 & 17:37 & 17:46 & X1.3 & 173.1 & NHJ & $\beta$$\gamma$$\delta$/$\beta$$\gamma$ \\
13234 & 25 & N24E63 & 2023/3/3 17:42 & 17:52 & 17:59 & X2.1 & 169.3 & NHJ & $\beta$$\gamma$$\delta$/$\beta$$\gamma$$\delta$ \\
13256 & 25 & S23E59 & 2023/3/29 2:18 & 2:33 & 2:40 & X1.2 & 345.8 & AJ & $\beta$$\gamma$ \\
13354 & 25 & N17E55 & 2023/7/2 22:54 & 23:14 & 23:58 & X1.1 & 169.6 & NHJ & $\beta$$\gamma$$\delta$/$\beta$$\gamma$$\delta$ \\
13386 & 25 & N18E59 & 2023/8/5 21:45 & 22:21 & 22:44 & X1.6 & 171.9 & NHJ & $\beta$/$\beta$ \\
13514 & 25 & N13E43 & 2023/12/14 16:47 & 17:02 & 17:12 & X2.8 & 226.6 & AJ & $\beta$ \\
13575 & 25 & S37E74 & 2024/2/9 12:53 & 13:14 & 13:32 & X3.3 & 146.8 & AHJ & $\beta$$\delta$/$\beta$$\gamma$$\delta$ \\
13614 & 25 & N19W08 & 2024/3/23 0:58 & 1:33 & 2:21 & X1.1 & 173.4 & NHJ & $\beta$/$\beta$ \\
13615 & 25 & S16W14 & 2024/3/23 0:58 & 1:33 & 2:21 & X1.1 & 11.3 & NHJ & $\beta$$\gamma$$\delta$/$\beta$$\gamma$$\delta$ \\
 & 25 & S16E64 & 2024/3/28 20:29 & 20:56 & 21:01 & X1.1 & 0.4 & NHJ & $\beta$$\gamma$$\delta$/$\beta$$\gamma$$\delta$ \\
13663 & 25 & N27E19 & 2024/5/5 11:41 & 11:54 & 12:16 & X1.2 & 161.9 & NHJ & $\beta$$\gamma$$\delta$/$\beta$$\gamma$$\delta$ \\
 & 25 & N27E52 & 2024/5/8 1:33 & 1:41 & 1:48 & X1.0 & 162.8 & NHJ & $\beta$$\gamma$$\delta$/$\beta$$\gamma$$\delta$ \\
 & 25 & N27W12 & 2024/5/3 2:11 & 2:22 & 2:27 & X1.7 & 159.9 & NHJ & $\beta$$\gamma$$\delta$/$\beta$$\gamma$$\delta$ \\
 & 25 & N27E16 & 2024/5/5 5:47 & 6:01 & 6:07 & X1.3 & 162.1 & NHJ & $\beta$$\gamma$$\delta$/$\beta$$\gamma$$\delta$ \\
 & 25 & N27E30 & 2024/5/6 5:38 & 6:35 & 6:47 & X4.5 & 163.2 & NHJ & $\beta$$\gamma$$\delta$/$\beta$$\gamma$$\delta$ \\
13664 & 25 & S19E30 & 2024/5/9 17:23 & 17:44 & 18:00 & X1.1 & 341.6 & AJ & $\beta$$\gamma$$\delta$ \\
 & 25 & S19E47 & 2024/5/11 1:10 & 1:23 & 1:39 & X5.8 & 347.1 & AJ & $\beta$$\gamma$$\delta$ \\
 & 25 & S19E20 & 2024/5/8 21:08 & 21:40 & 23:10 & X1.0 & 324.6 & AJ & $\beta$$\gamma$$\delta$ \\
 & 25 & S19E09 & 2024/5/8 4:37 & 5:09 & 5:32 & X1.0 & 310.0 & AJ & $\beta$$\gamma$$\delta$ \\
 & 25 & S19E37 & 2024/5/10 6:27 & 6:54 & 7:06 & X3.9 & 345.4 & AJ & $\beta$$\gamma$$\delta$ \\
 & 25 & S19E25 & 2024/5/9 8:45 & 9:13 & 9:36 & X2.2 & 336.5 & AJ & $\beta$$\gamma$$\delta$ \\
13738 & 25 & S15E55 & 2024/7/14 2:23 & 2:34 & 2:48 & X1.2 & 18.5 & NHJ & $\beta$$\gamma$$\delta$/$\beta$$\gamma$ \\
13764 & 25 & S04W01 & 2024/7/29 2:33 & 2:37 & 2:43 & X1.5 & 358.9 & AJ & $\beta$/$\alpha$ \\
13777 & 25 & S09E26 & 2024/8/8 19:01 & 19:35 & 19:57 & X1.3 & 330.8 & AJ & $\beta$$\gamma$$\delta$/$\beta$$\gamma$ \\
13784 & 25 & N16W06 & 2024/8/14 6:00 & 6:40 & 7:08 & X1.1 & 197.3 & AJ & $\beta$$\gamma$$\delta$/$\beta$$\delta$ \\
13842 & 25 & S11E35 & 2024/10/3 12:08 & 12:18 & 12:27 & X9.0 & 332.5 & AJ & $\beta$$\gamma$$\delta$ \\
 & 25 & S11E88 & 2024/10/7 19:02 & 19:13 & 19:31 & X2.1 & 342.7 & AJ & $\beta$$\gamma$$\delta$ \\
 & 25 & S11E87 & 2024/10/7 20:03 & 20:59 & 21:27 & X1.0 & 342.1 & AJ & $\beta$$\gamma$$\delta$ \\
 & 25 & S11E13 & 2024/10/1 21:58 & 22:20 & 22:29 & X7.1 & 326.4 & AJ & $\beta$$\gamma$$\delta$ \\
13848 & 25 & N14E07 & 2024/10/9 1:25 & 1:56 & 2:43 & X1.8 & 221.4 & AJ & $\beta$$\gamma$$\delta$ \\
13883 & 25 & S08W18 & 2024/11/6 13:24 & 13:40 & 13:46 & X2.3 & 6.3 & NHJ & $\beta$$\gamma$$\delta$/$\gamma$$\delta$ \\
13912 & 25 & S10E40 & 2024/12/8 8:50 & 9:06 & 9:10 & X2.2 & 357.7 & AJ & $\beta$$\gamma$/$\beta$ \\
13936 & 25 & N16E14 & 2024/12/29 7:08 & 7:18 & 7:34 & X1.1 & 177.1 & NHJ & $\beta$$\delta$/$\beta$$\delta$ \\
14098 & 25 & S05W44 & 2025/5/25 1:46 & 1:52 & 1:57 & X1.1 & 342.1 & AJ & $\beta$/$\beta$ \\
14114 & 25 & N18W10 & 2025/6/17 21:38 & 21:49 & 21:54 & X1.2 & 150.0 & NHJ & $\beta$$\gamma$$\delta$/$\beta$$\gamma$$\delta$ \\
14274 & 25 & N23W23 & 2025/11/11 9:49 & 10:04 & 10:22 & X5.2 & 244.3 & AJ & $\beta$$\gamma$$\delta$/$\beta$$\gamma$$\delta$ \\
 & 25 & N21W58 & 2025/11/14 7:44 & 7:30 & 7:40 & X4.0 & 248.3 & AJ & $\beta$$\gamma$$\delta$/$\beta$$\gamma$$\delta$ \\
 & 25 & N23E03 & 2025/11/9 7:01 & 7:35 & 7:52 & X1.8 & 231.4 & AJ & $\beta$$\gamma$$\delta$/$\beta$$\gamma$$\delta$ \\
 & 25 & N23W14 & 2025/11/10 8:55 & 9:19 & 9:36 & X1.2 & 238.5 & AJ & $\beta$$\gamma$$\delta$/$\beta$$\gamma$$\delta$ \\
\hline
\end{tabular}
\end{table}

\subsection{Tilt Calculation and Quadrant Definition}
\label{sec:magnetic_tilt_definition}

The magnetic tilt angle $\theta$ of a sunspot group is defined as the orientation of the vector connecting the centroid of positive magnetic polarity to that of negative magnetic polarity, measured relative to the solar equatorial direction (eastward) within the heliographic coordinate system. Its valid angular range spans from $0^{\circ}$ to $360^{\circ}$.

\begin{figure}
\includegraphics[width=\linewidth]{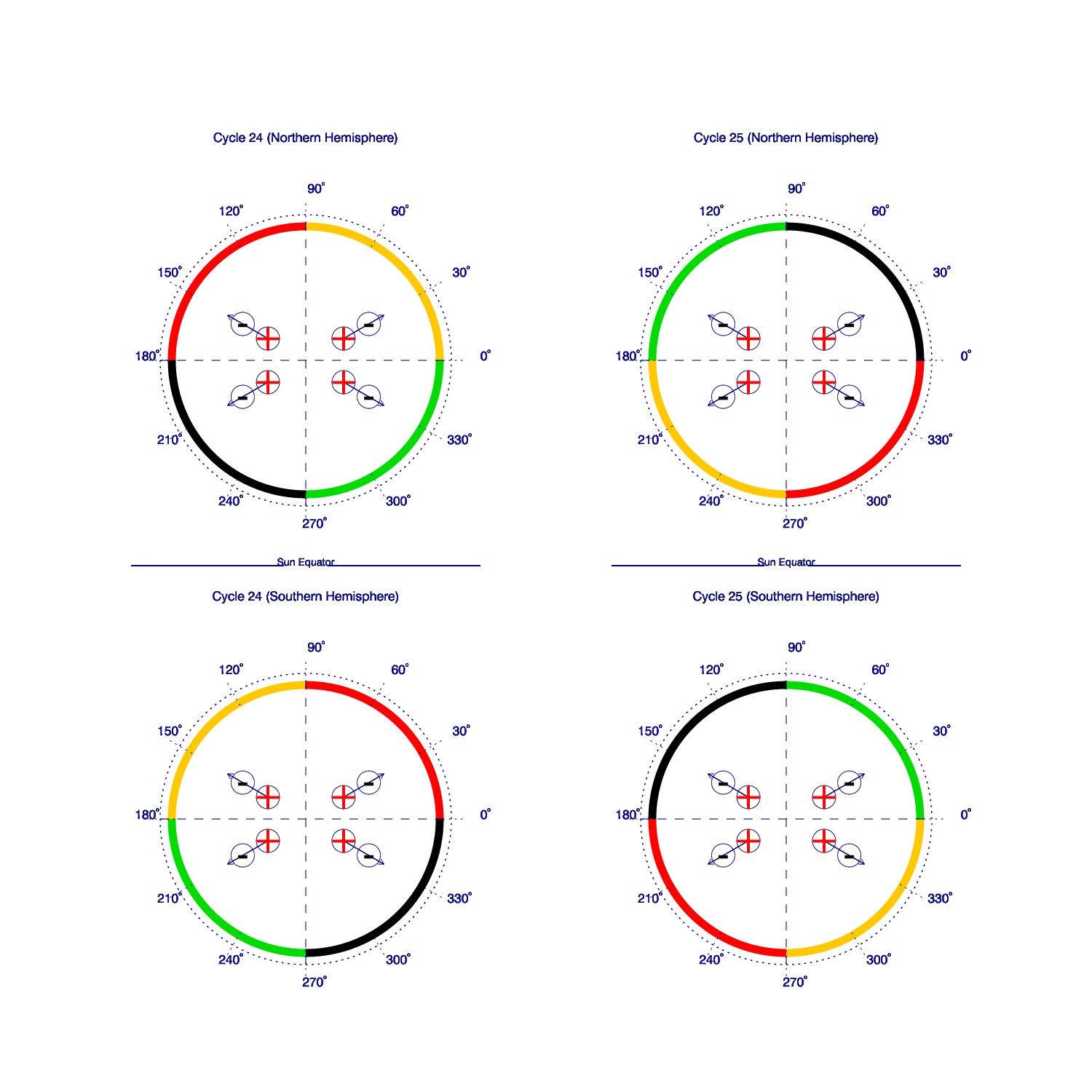}
\caption{The definition of four color-coded quadrants for magnetic tilts, in both northern and southern hemispheres and in both solar cycles 24 and 25. The horizontal line indicates the solar equator, north is upward.  Plus and minus signs ($+$/$-$) denote the positive and negative polarity sunspots. The vector line with arrow indicates the magnetic tilt angle. The four color-coded quadrants with the tilts range from 0$^\circ$ to 360$^\circ$ represent different tilt characters: normal Hale-Joy (NHJ: green), anti-Joy (AJ: yellow), anti-Hale (AH: red), anti-Hale and anti-Joy (AHJ: black).}
\label{combined_tilt_angle_polar}
\end{figure}

Based on established empirical knowledge of sunspot group tilt angles and the fundamental physical laws, we classify sunspot groups into four distinct quadrants (covering the full 0°–360° tilt-angle range) within a polar coordinate system where the angle increases counterclockwise, with 0° indicating the eastward direction. Separate classifications are applied to the northern and southern hemispheres, as illustrated in Figure \ref{combined_tilt_angle_polar} for Solar Cycles 24 and 25:

\begin{itemize}
\item Green quadrant: Normal Hale–Joy compliant configurations (NHJ)
\item Yellow quadrant: Anti-Joy characteristics (AJ)
\item Red quadrant: Anti-Hale properties (AH)
\item Black quadrant: Both anti-Hale and anti-Joy behaviour (AHJ)
\end{itemize}

In the polar representation shown in Figure \ref{combined_tilt_angle_polar}, the solar equator is represented by a horizontal line, with the positive direction oriented eastward (to the right, corresponding to 0° in the counterclockwise direction). Each panel is labelled at the top with the corresponding solar cycle number (24 or 25) and hemisphere (south or north). Plus and minus signs ($+$/$-$) indicate the positions of positive- and negative-polarity sunspots, respectively. Vector-direction lines with arrows represent the magnetic tilt angles of the sunspot groups.

The magnetic tilt angle in this study is determined through the following procedure using Space-weather HMI Active Region Patches (SHARP) magnetograms which have been remapped to the Cylindrical Equal-Area (CEA) coordinate system:
\begin{enumerate}
\item Centroid Calculation: The flux-weighted centroids for each polarity are computed using the formula:
\begin{equation}
\mathbf{C}_{\pm} = \frac{\sum B_{\pm}(\mathbf{r}) \cdot \mathbf{r}}{\sum B_{\pm}(\mathbf{r})} ~~~,
\end{equation}
\noindent where $B_{\pm}(\mathbf{r})$ represents the radial component of the magnetic field at position $\mathbf{r}$ for positive and negative polarities, respectively.

\item Magnetic Tilt Determination: The magnetic tilt $\theta$ is calculated as the azimuthal direction of the vector from $\mathbf{C}_{+}$ to $\mathbf{C}_{-}$ in the heliographic coordinate system.
\end{enumerate}

For each active region, we calculate the magnetic tilt angle using every available magnetogram during its passage across the solar disk.  This yields a time series of typically several hundred to over one thousand tilt 
angle measurements (given the 12-minute cadence of HMI magnetograms) for each active region. The tilt angle at each individual time step, together with the hemisphere and solar cycle information, is then compared with the classification scheme 
presented in Figure \ref{combined_tilt_angle_polar} to determine whether the active region at that specific moment complies with or deviates from Hale's and/or Joy's laws.

%%%%%%%%%%%%%%%

\section{Analysis and Results}
\label{sec:results} 

\subsection{Two Active Regions as an Example}

\begin{figure}
\includegraphics[width=\linewidth]{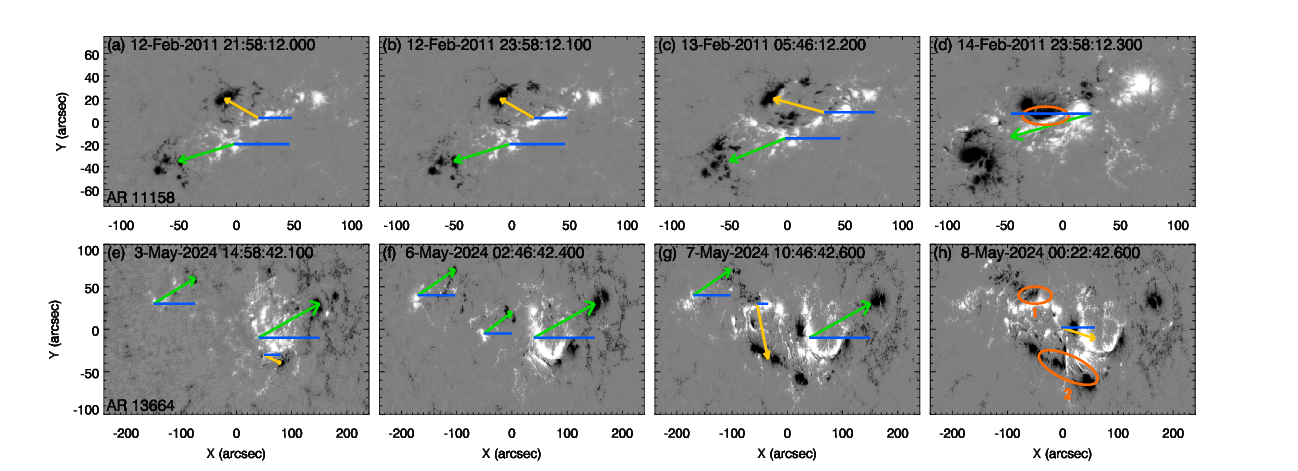}
\caption{Magnetic field configuration and tilt analysis of NOAA 11158 and NOAA 13664. Panels (a)-(c) depict tilts of regional magnetic features, whereas panel (d) shows the global tilt of the entire active region, for NOAA 11158. Panels (e)-(h) are for NOAA 13664. The grayscale background image in each panel shows the radial magnetic field distribution. The blue line outlines the solar equator direction (east to the right). Green and yellow arrows represent the magnetic tilt arrows (from positive to negative polarity). Green means that Joy's law is followed, yellow means that Joy's law is violated. Orange ellipses mark the key regions of opposite-polarity collisions.}
\label{fig:11158tilt}
\end{figure}

Figure \ref{fig:11158tilt} presents two example ARs, of their magnetic field configurations and tilt angle characteristics, one for NOAA 11158 in panels (a)-(d), one for NOAA 13664 in panels (e)-(h). 
AR 11158 belongs to solar cycle 24 and is located in the southern hemisphere. NOAA 13664 belongs to solar cycle 25 and is also located in the southern hemisphere. 

For NOAA 11158, it is classified as a normal active region (NHJ) overall, as shown by the green tilt arrow calculated from the whole active region in panel (d). However, during its formation stage, this AR exhibited a more complex evolution, marked by the interaction between a normal emerging flux system (indicated by the green arrows) and an anti‑Joy emerging flux system (indicated by the yellow arrows) in panels (a)–(c). Observations show that the leading positive polarity of the normal emerging region collided with the trailing negative polarity of the anti‑Joy region as shown by the orange circle in panel (d). This interaction generated shear motions (the relative motion of opposite polarities along the polarity inversion line) and facilitated the accumulation of magnetic energy within the active region.

NOAA 13664 is categorized overall as an anti-Joy active region (AJ) as shown in panel (h) by the yellow tilt arrow. Panel (f) shows that in its early developmental stage the region presents three separate normal emergence regions. However, subsequently  the central emergence zone underwent a transition to exhibit anti-Joy emergence characteristics [panels (f-g)]. This structural transformation triggered two distinct sets of polar collisions: the trailing positive polarity sunspots clashed with the leading negative polarity sunspots of the left adjacent region (orange circle 1), while the leading negative polarity sunspots interacted with the trailing positive polarity sunspots of the right adjacent region (orange circle 2). Collectively, these collisions generated shear across the three emergence regions and led to the buildup of magnetic energy.
Here in this figure we see that the coexistence and interaction of emergence regions with opposing magnetic tilts (i.e., magnetic structures with opposite writhe) within active regions tend to induce collisions between positive and negative magnetic polarities, thereby promoting shear deformation and the accumulation of magnetic energy.

\begin{figure}
\includegraphics[width=0.8\linewidth]{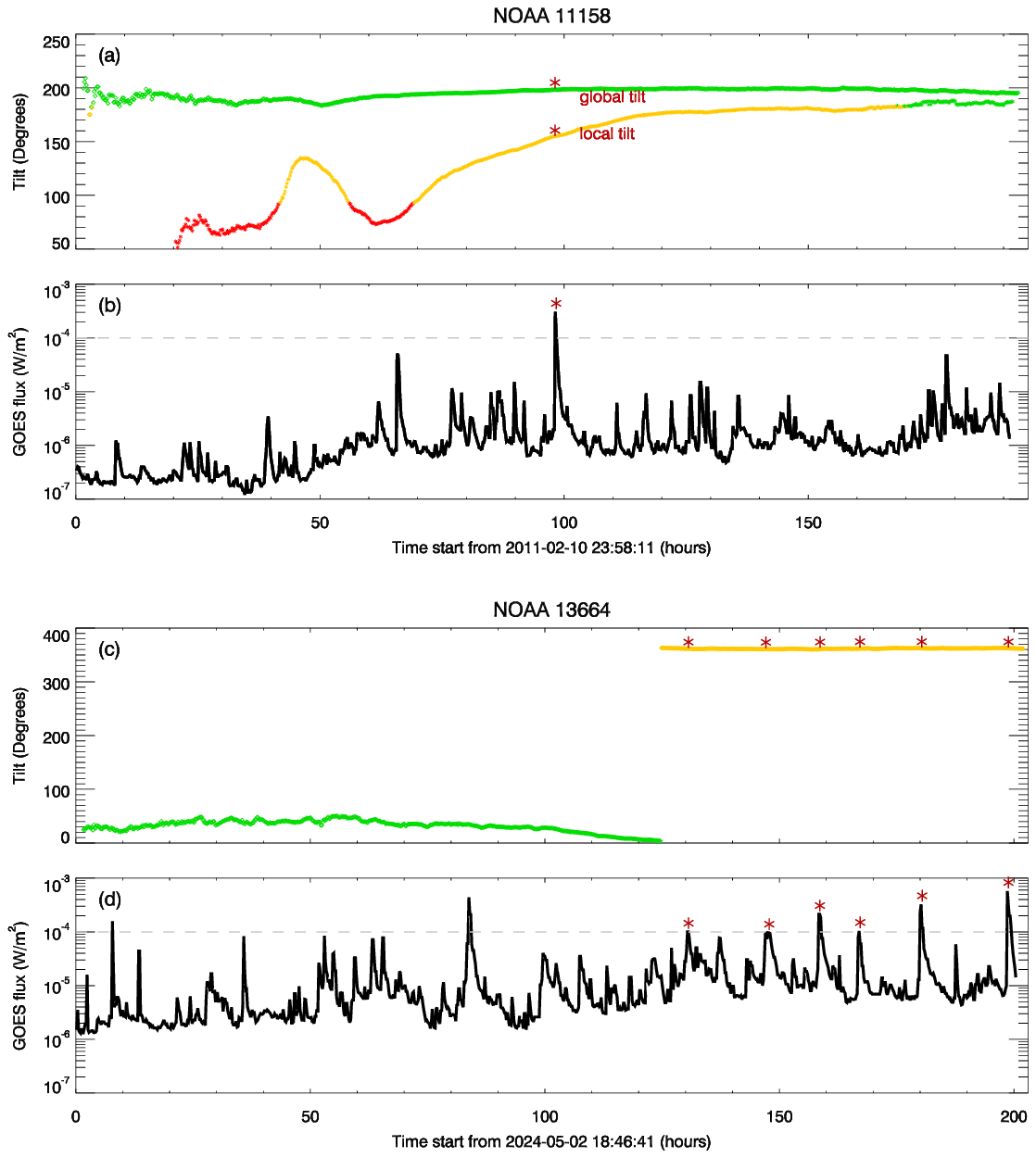}
\caption{(a, c) Temporal evolution of magnetic tilt angles for NOAA 11158 (a) and NOAA 13664 (c). For NOAA 11158, both the global tilt (labeled "global") and a local anti-Joy tilt (labeled "local") are shown. For NOAA 13664, only the global tilt is displayed. Tilt angles are color-coded according to the quadrant classification in Figure \ref{combined_tilt_angle_polar}, with red asterisks marking X-class flare times. (b, d) Corresponding GOES soft X-ray (1-8~\AA) flux for NOAA 11158 (b) and NOAA 13664 (d); thin gray dashed lines indicate the X1.0 level at $1\times10^{-4}$~W~m$^{-2}$.}
\label{titlegoes158664_4x1}
\end{figure}

These complex, multi-polarity interactions within an active region are reflected in its overall magnetic configuration and can be quantitatively characterized by the global magnetic tilt angle. The temporal evolution of this integrated parameter thus provides us information on the magnetic field configuration state of the active region together with its potential for magnetic energy release.
This is more evident in Figure \ref{titlegoes158664_4x1} where the temporal evolution of the global magnetic tilt, alongside GOES X-ray flux, are shown. The tilt measurements are color-coded according to their quadrant classification, with red asterisks marking the occurrence times of X-class flares on both the tilt and GOES flux profiles. In the GOES flux panels, gray dashed lines indicate the X1.0 level at 1$\times$10$^{-4}$ W m$^{-2}$. Here we observe a clear association of flare productivity with magnetic tilt quadrant. NOAA 11158, which maintained a green-coded tilt (labeled as "global" in the curve, NHJ: indicating conformity with both Hale and Joy laws) throughout its evolution, produced only one X-class flare. In contrast, all six X-class flares from NOAA 13664 occurred when its global tilt was in the yellow quadrant (AJ), revealing consistent anti-Joy characteristics during these eruptive episodes. For AR 13664, the two X-class flares shown without asterisks belong to a different active region, AR 13663.

We also examined the local tilt evolution of NOAA 11158. As shown by the yellow arrows in panels (c) and (d) of Figure \ref{fig:11158tilt}, an anti-Joy emerging flux system was present locally. The temporal evolution of this local tilt (shown as the local curve in Figure \ref{titlegoes158664_4x1}) exhibits both red (AH) and yellow (AJ) signatures, with a transition between different quadrants during the early emergence phase. In contrast, once the local system matured, the tilt stabilized in the AJ (yellow) quadrant and remained thereafter on the whole. A clear association with flare occurrence then emerged: with the increasing trend of the local AJ tilt during this mature phase, X-class flare began to erupt. This further demonstrates that local tilt anomalies, once reliably established, are closely correlated with major flare activity.

\subsection{Local Tilt in NHJ Active Regions}

\begin{figure}
\includegraphics[width=\linewidth]{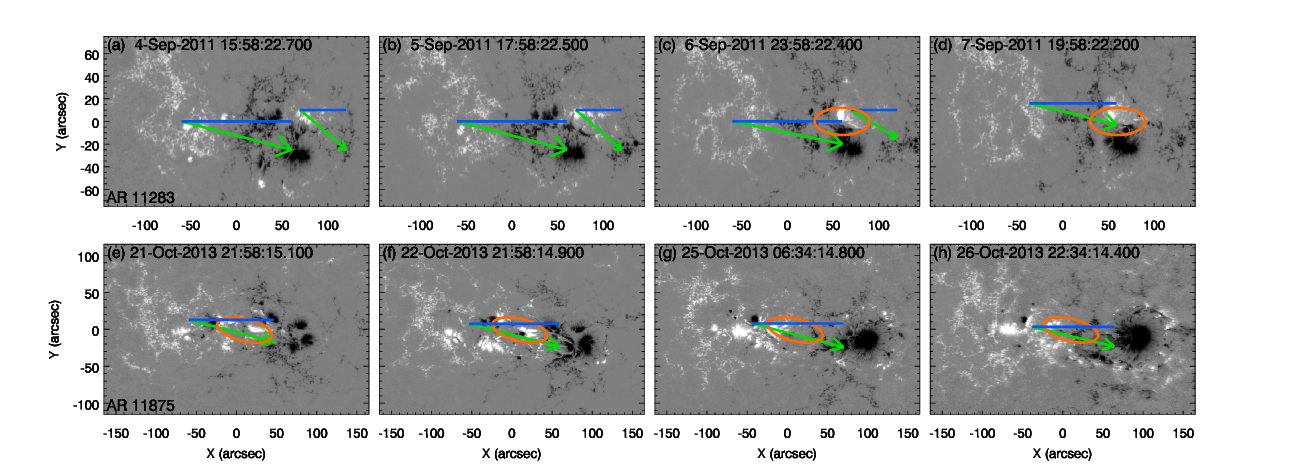}
\caption{Magnetic field configuration and local tilt analysis of NOAA 11283 and NOAA 11875. Panels (a)-(c) depict tilts of regional magnetic features, whereas panel (d) shows the global tilt of the entire active region for NOAA 11283. Panels (e)-(h) show the global tilt evolution for NOAA 11875. The grayscale background image in each panel shows the radial magnetic field distribution. The blue line outlines the solar equator direction (east to the right). Green arrows represent magnetic tilt arrows (from positive to negative polarity), indicating that Joy's law is followed. Orange ellipses mark key regions of opposite-polarity collisions, where shear motions are present.}
\label{fig:11283tilt}
\end{figure}

We further examined the local tilt evolution for all globally normal (NHJ) active regions in our sample. Among the 18 NHJ regions, we identified three categories based on their local tilt characteristics:

\begin{enumerate}
\item Regions with clear local anomalous tilt (12 out of 18): As shown in Figure \ref{fig:11158tilt} for NOAA 11158, local anti-Joy tilt features (yellow arrows) coexisted with the global NHJ configuration. The temporal evolution of these local anomalous tilts showed a clear association with X-class flare occurrence.

\item Regions with local normal tilt (2 out of 18): As shown in Figure \ref{fig:11283tilt} panels (a)-(c) for NOAA 11283, the local tilt features were normal (green arrows), but strong shear motions (orange ellipses), i.e., the squeezing and relative motion of opposite polarities, were present between opposite polarities from different bipoles. The global tilt remained in the NHJ quadrant (panel d).

\item Regions with complex local configurations (4 out of 18): As shown in Figure \ref{fig:11283tilt} panels (e)-(h) for NOAA 11875, the local tilt features were too complex to be unambiguously classified as either normal or anomalous. However, clear shear motions (orange ellipses) were present between opposite polarities, and the global tilt remained in the NHJ quadrant.
\end{enumerate}

Despite the diversity of local tilt characteristics among NHJ regions, a common feature is that all 18 NHJ regions in our sample harbor localized magnetic structures with strong shear motions — i.e., relative motion of opposite polarities — in the vicinity of the X-class flare locations. This universal presence of local shear, whether associated with normal or anomalous tilts, further confirms that X-class flares are consistently associated with abnormal magnetic configurations at either the global or local scale, ultimately manifesting as complex sunspot configurations (as listed in Table~\ref{tab:flare_summary}).

\subsection{The Statistics for Global Tilt}

\begin{figure}
\includegraphics[width=\linewidth]{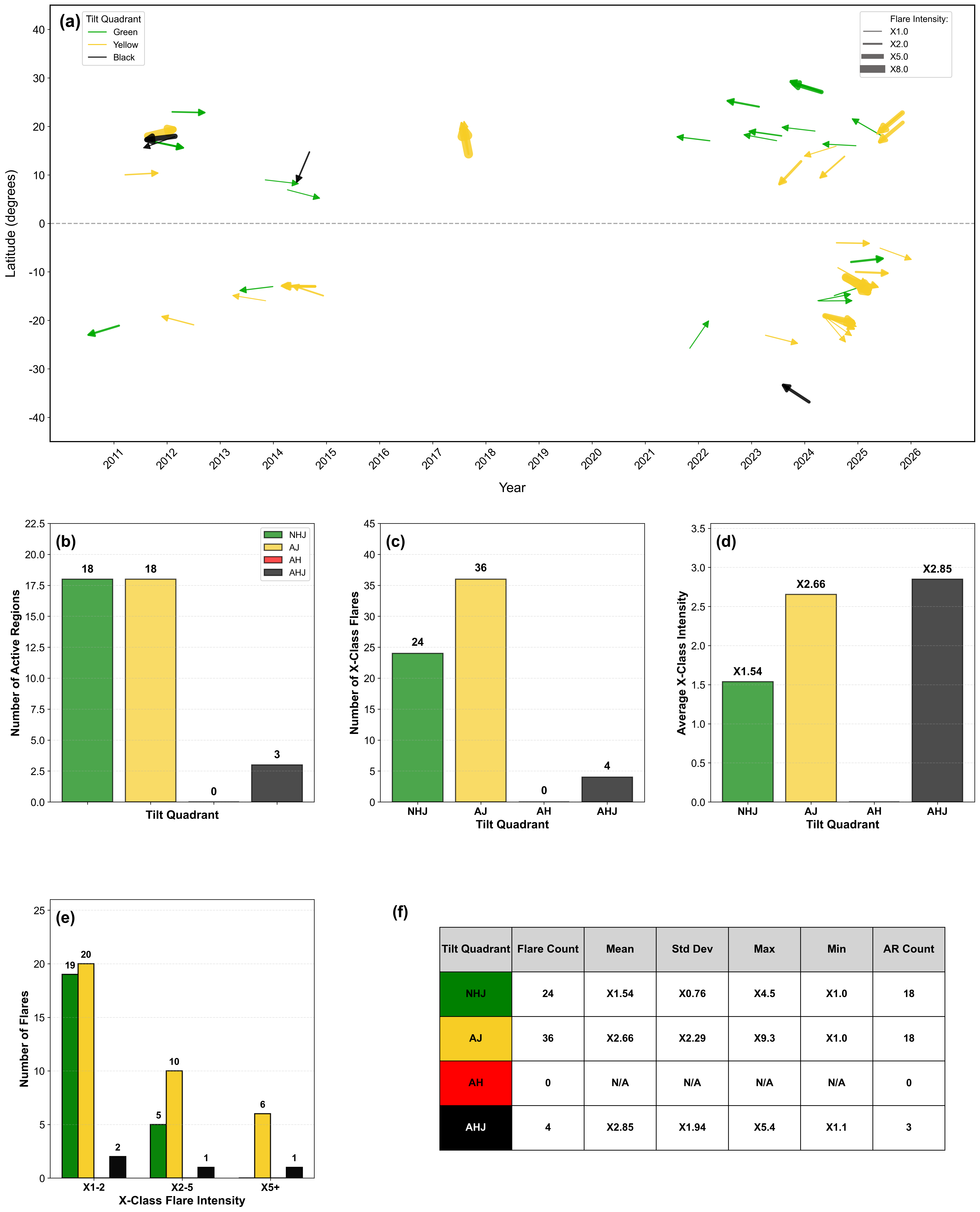}
\caption{
(a) Time-latitude plot of magnetic tilt angles for each of X-class flare listed in Table~\ref{tab:flare_summary}. Arrows outline the tilt angles and colored by magnetic configuration: green (NHJ), yellow (AJ), and black (AHJ). Arrow thickness scales with flare class.
(b) The number of active regions in each quadrant.
(c) The total number of flares in each quadrant.
(d) The mean flare class for each quadrant.
(e) The number of flares across different intensity classes (X1-2, X2-5, X5+).
(f) A statistical summary table of the flare analysis.
}
\label{xclass_statistics}
\end{figure}

We extended the above analysis of global tilt to our full sample. We compiled the flare-associated tilt quadrants for all events and the results are listed in the ninth column of Table \ref{tab:flare_summary}. These data are also visualized in panel (a) of Figure \ref{xclass_statistics}.

In this panel, the horizontal axis represents time and the vertical axis represents latitude. Each arrow represents the magnetic tilt angle at the time of an X-class flare, following the convention that 0$^\circ$ points eastward and increases counterclockwise to 360$^\circ$. The color of each arrow corresponds to its Hale-Joy quadrant classification (green: NHJ, yellow: AJ, black: AHJ), and the arrow thickness scales with the flare intensity.
Figure \ref{xclass_statistics}(a) also exhibits an approximate butterfly diagram pattern for X-class flares: during solar maximum, flares tend to occur at higher latitudes; near minimum, they are confined to lower latitudes and become fewer. Due to the long time span (2008–2025), some arrows at similar latitudes appear densely packed or overlapping; however, each arrow represents a distinct X-class flare. Importantly, no active region in our sample produced another X-class flare after reappearing in a subsequent Carrington rotation.
The table and this visualization together provide a clear mapping of flare-producing active regions to their global magnetic tilt quadrants, forming the observational basis for our statistical analysis of the relationship between magnetic tilt configuration and flare productivity.

The statistical results of our sample are presented in panels (b)–(f) of Figure \ref{xclass_statistics}, which shows a clear relationship between the magnetic tilt configuration and X-class flare productivity. 
The histogram in the panel (b) shows the distribution of active regions by magnetic classification: 18 active regions had tilts in the green (NHJ) quadrant, 18 active regions belonged to the AJ (yellow) quadrant, and only 3 active regions were classified as AHJ in the black quadrant. This distribution is complemented by the panel (c), which displays the total flare counts: 24 flares originated from active regions in the NHJ quadrant, 36 from active regions in the AJ quadrant, and 4 from AHJ active regions. The higher flare count in the AJ quadrant suggests that the regions with this tilt configuration tend to be more prolific in producing X-class flares.

A clear pattern emerges from the panel (d) when comparing the mean flare class for each quadrant. For active regions in the NHJ quadrant, the mean flare class is X1.54, whereas for AJ and AHJ active regions, the mean flare classes are  X2.66 and X2.85 respectively, substantially higher than that of active regions in the NHJ quadrant.

The panel (e) reveals the flare intensity distribution across different classes. For X1-2 class flares, 19 are from active regions in the NHJ quadrant, 20 are from active regions in the AJ quadrant, and 2 from AHJ active regions. The distribution shifts notably in the X2-5 intensity range, with 5 flares from NHJ active regions, 10 from AJ active regions, and 1 from AHJ active regions. Most significantly, the X5+ category contains 6 flares  from AJ active regions exclusively and 1 from a AHJ active region, with none from NHJ active regions.

Although the number of NHJ active regions (18) is comparable to the total number of AJ and AHJ active regions (21), our analysis reveals a stark contrast in their energetic outputs. The AJ and AHJ active regions show significantly higher flare productivity, both in terms of occurrence frequency and flare class. This pattern is further corroborated by the statistical summary in the panel (f), where the maximum recorded flare classes distinctly indicate the predisposition of AJ and AHJ magnetic configurations for producing major flares. Notably, the complete absence of AH active regions in our sample suggests that such magnetic tilt configurations are less favorable to occur alone for flare production. 
%?Notably, the complete absence of AH active regions in our sample suggests that such magnetic configurations represent an uncommon situation. 

%%%%%%%%%%%%%%%%%%

\section{Discussion and Conclusion}
\label{sec:disc}

\subsection{Discussion}

The statistical correlations identified in our study provide interesting implications on a few physical mechanisms that link anomalous magnetic tilt to enhanced flare productivity and flare intensity.

Active regions with Anti-Joy (AJ) magnetic tilt configurations naturally develop strong magnetic shear. Here, shear refers to the relative motion of opposite polarities along the polarity inversion line (e.g., the orange circles in Figure \ref{fig:11158tilt}). Such relative motions stretch and deform the magnetic field lines, favoring the shearing and upward expansion of field lines, facilitating the accumulation of non-potential energy \citep{2014masu.book.....P} and potentially lowering the threshold for eruptive instabilities \citep{2006ApJ...644..575Z}.
In contrast, major flares in globally normal (NHJ) regions are primarily powered by intense, localized magnetic anomalies (e.g., satellite sunspots) embedded within an otherwise compliant large-scale configuration. 
This dichotomy gives rise to the concept of opposite magnetic helicity collision, wherein magnetic structures with opposite writhe (as indicated by tilt angle) interact. Such interactions effectively concentrate energy, thereby creating conditions conducive to major flares.

These two flaring modes exhibit distinct degrees of ``configurational conflict" with the ambient solar magnetic field. In globally anomalous (AJ/AHJ) regions, this conflict is fundamental and large-scale, fostering highly asymmetric polarity distributions conducive to magnetic collisions and energy buildup. In contrast, for globally normal (NHJ) regions, the conflict is confined to localized anomalies (e.g., satellite sunspots) where contrasting helicity can concentrate energy.

Therefore, our quadrant classification systematically captures these distinct physical states by quantifying the degree of configurational conflict and, consequently, the capacity for storing magnetic free energy. This framework directly explains the observed intensity hierarchy (NHJ → AJ → AHJ), where progressively more anomalous configurations correlate with both higher stored energy and the potential for more intense flares.

The role of local tilt anomalies is particularly important for understanding flare activity in globally normal regions. As shown in Figure \ref{titlegoes158664_4x1} for NOAA 11158, the local anti-Joy tilt (curve labeled "local") exhibited a clear temporal association with X-class flare occurrence, even though the global tilt remained in the NHJ quadrant. We have examined all 18 globally normal (NHJ) regions in our sample and classified their local tilt characteristics as follows: 12 regions exhibit clear local anomalous tilt; 2 regions exhibit local normal tilt but with strong shear motions; and the remaining 4 regions have complex configurations that preclude unambiguous classification. This demonstrates that local tilt anomalies serve as key indicators of magnetic non-potentiality and flare productivity, and that X-class flares are universally associated with abnormal tilt configurations — whether at the global scale (AJ/AHJ) or the local scale (within NHJ regions).

The absence of pure AH (red quadrant) regions in our X-class flare sample suggests that such configurations may be generally uncommon, even among non-flaring active regions. When Hale's law is violated, a deviation from Joy's law often accompanies it, leading to an AHJ classification rather than pure AH. This implies that AH configurations frequently exhibit leading polarities tilted toward the pole, and such tilts tend to co-occur with anti-Joy characteristics. This explains why we have AHJ regions in our sample but no pure AH regions.

\subsection{Conclusion}

Our systematic analysis of 64 X-class flares from 39 active regions identifies a clear correlation between the anomalous magnetic tilt configuration, i.e.,  the violation of Hale’s and/or Joy’s laws, and the increased flare productivity as well as the increased flare intensity. Key findings from this analysis are as follows:

\begin{enumerate}
\item Two Distinct Types of Flare-productive Regions: 

Flare-productive active regions can be divided into two categories: Over half (54\%, 21 out of 39) of the active regions are globally anomalous, showing systematic violations of Hale's and/or Joy's laws at the global scale (AJ or AHJ). The remaining 46\% (18 out of 39) are globally normal (NHJ) regions that comply with both laws in their global magnetic tilt. Notably, even though these 18 regions are classified as normal based on their overall tilt, all of them harbored localized magnetic features (e.g., satellite sunspots) with strong shear motions. Among these, 12 exhibited clear local anomalous (anti-Joy) tilts, 2 showed local normal tilts but with strong shear, and the remaining 4 had complex configurations where unambiguous classification was not possible.

%\item Globally Anomalous Regions (53\%=(16+3)/36): Over half of flare-productive regions exhibit globally anomalous tilt characters (AJ or AHJ). While only about 40\% of all active regions exhibit anomalous tilt configurations (irrespective of flare productivity), this fraction rises to 53\% among those producing X-class flares, underscoring the preferential flare productivity of anomalous regions. %From ppt jiangjie where is citep

\item Enhanced Flare Productivity:
 
Anti-Joy (AJ) regions exhibit significantly greater flare productivity, with an average of 2.00 flares per active region (36/18)—a 50\% increase compared to that of the normal Hale-Joy (NHJ) regions, which only yield 1.33 flares per active region (24/18).

\item Flare Intensity Gradation Across Magnetic Configurations:

A distinct intensity gradient of X-class flares is observed across different tilt quadrants: X1–2 class flares occur in both normal and anomalous quadrant types (i.e., NHJ, AJ, and AHJ); X2–5 class flares are dominated by AJ regions; and X5+ class flares are restricted exclusively to the AJ and AHJ quadrants. This flare intensity trend is further confirmed by the mean flare classes of each quadrant: X1.54 (NHJ), X2.66 (AJ), and X2.85 (AHJ).

\item Absence of Pure Anti-Hale Regions: 

Pure AH configurations are absent from our X-class flare sample, indicating that anti-Hale characteristics typically co-occur with anti-Joy traits (i.e., AHJ). This makes AJ the dominant anomalous factor, while AHJ configurations are relatively rare. Pure AH regions hardly appear among X-class flare producers.
\end{enumerate}

In summary, deviations from Hale’s and Joy’s laws— globally or just locally—act as reliable indicators of magnetic non-potentiality in active regions. Such deviations are significantly correlated with both a higher occurrence rate and greater intensity of X-class solar flares, confirming a direct link between anomalous magnetic tilt configurations and enhanced flare activity.

\begin{acknowledgments}
%We sincerely thank the anonymous referee for the valuable suggestions and constructive comments, which have significantly improved the overall quality of this paper.
This research was supported by the National Key R\&D Program of China (No. 2022YFF0503001, 2021YFA1600500), National Natural Science Foundation of China (grant Nos. 12273059 and 12473053),  China's Space Origins Exploration Program (GJ11020400), the Strategic Priority Research Program on Space Science of the Chinese Academy of Sciences (Grant No. XDA15320000), and the specialized research fund for State Key Laboratory of Solar Activity and Space Weather.
\end{acknowledgments}

\begin{contribution}
%%This section gives authors the space to recognize author contributions. The text inside this environment is NOT counted towards the total word quanta. At a minimum, manuscripts are expected to include this text:
All authors contributed equally to the paper.
\end{contribution}

\bibliography{lius}{}
\bibliographystyle{aasjournalv7}

\end{document}